\documentclass[12pt]{iopart}

\usepackage{iopams,harvard}
\usepackage{graphicx}
\usepackage{dcolumn}
\usepackage{psfrag}
\usepackage{float}
\newcommand{\ve}[1]{\ensuremath{\mbox{\boldmath$#1$}}}
\newcommand{\ma}[1]{\ensuremath{\mathbb{#1}}}
\newcommand{\st}{{\rm St}}
\newcommand{\ku}{{\rm Ku}}
\renewcommand{\tr}{{\rm Tr}}
\usepackage[dvips,dvipsnames,cmyk]{xcolor}
\usepackage{color}
\begin{document}

\title[Inertial-particle dynamics in turbulent flows]{Inertial-particle dynamics in turbulent flows: caustics, concentration
fluctuations, and random uncorrelated motion}

\author{K. Gustavsson$^{1)}$, E. Meneguz$^{2,3)}$, M. Reeks$^{2)}$,
and B. Mehlig$^{1)}$}

\address{$^{1)}$Department of Physics, Gothenburg University, 41296
Gothenburg, Sweden\\
$^{2)}$School of Mechanical \& Systems Engineering, Newcastle
University, Newcastle NE1 7RU, UK\\
$^{3)}$Met Office, Exeter EX13PB, UK
}
\begin{abstract}
We have performed numerical simulations of inertial particles
in random model flows in the white-noise limit
(at zero Kubo number, $\ku=0$)
and at finite Kubo numbers. Our results for the moments
of relative inertial particle velocities are in good agreement
with recent theoretical results \cite{Gus10} based on the 
formation of phase-space singularities in the inertial
particle dynamics (caustics). We discuss the relation
between three recent approaches describing
the dynamics and spatial distribution of inertial particles
suspended in turbulent flows:
caustic formation,
real-space singularities of the deformation tensor,
and random uncorrelated motion. We discuss how the phase- and real-space
singularities are related. Their formation is well understood in terms
of a local theory. We discuss implications for random uncorrelated
motion. 
\end{abstract}
\maketitle

\section{Introduction}

The dynamics of particles suspended in randomly mixing or turbulent
flows (\lq turbulent aerosols') has been studied intensively for
several decades.
Recently, substantial progress
in understanding the dynamics of turbulent aerosols has been achieved
(see the papers published in this special issue and the references cited therein).

The phenomenon of spatial clustering of independent point particles
subject to Stokes drag in turbulent flows is now well understood:
below the dissipative length scale (where the fluid flow is smooth)
the particles eventually cluster onto a fractal set in configuration
space. The corresponding fractal dimension has been determined by means of direct
numerical simulations \cite{Bec03b} as well as theoretical approaches
\cite{Wil07,Gus11}. Different mechanisms [\lq preferential
concentration' \cite{Max87} and \lq multiplicative amplification'
\cite{Wil07,Gus11}] contribute to spatial clustering.
A third mechanism giving rise to particle clustering
was recently studied by
following the deformation of an infinitesimally small volume
of particles transported along a particle trajectory
[\lq full Lagrangian method' \cite{Ijz10}].
The small volume may vanish at isolated singular points in time, giving  rise to instantaneous singularities
in the particle-concentration field. Using this approach the statistical properties of these singularities
were analysed by Meneguz \harvardand\ Reeks (2011).

One important reason for studying spatial clustering of inertial particles
is that this phenomenon is argued to enhance the rate at which collisions
occur in turbulent aerosols at small values of the \lq Stokes number'.
This dimensionless parameter, ${\rm St}=(\gamma\tau)^{-1}$, is given
in terms of the particle damping rate $\gamma$ and the relevant correlation
time $\tau$ of the flow. Both are defined below.
\begin{figure}[t]
\hfill\includegraphics[width=10cm]{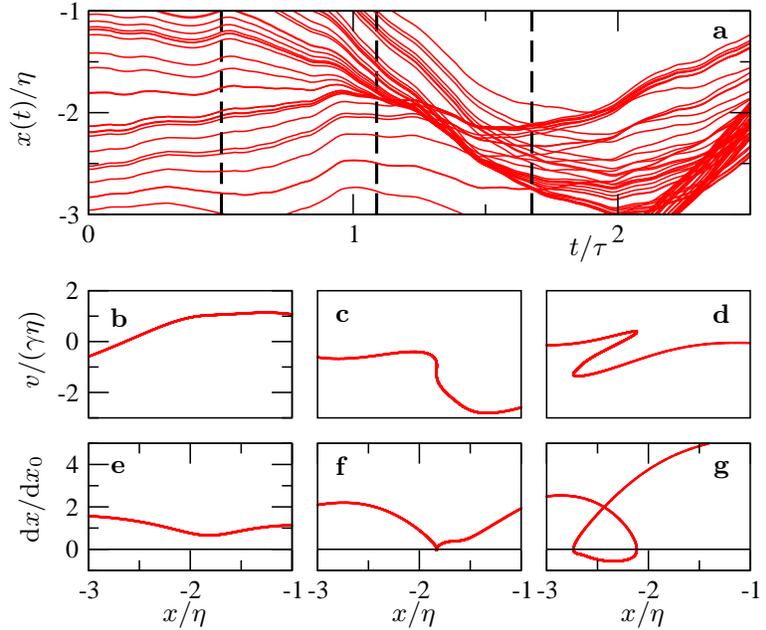}
\caption{\label{fig:1} {\bf a} Trajectories of a one-dimensional model, particle positions
as a function of time. Also shown ({\bf b}, {\bf c}, {\bf d}):
phase-space manifolds (velocity $v$ versus position $x$) demonstrating how the phase-space manifold
folds over at a caustic. 
Panels {\bf a} to {\bf d} are similar to
Fig.~1 in Gustavsson \harvardand\ Mehlig (2011{\em a}).
Also shown ({\bf e}, {\bf f}, {\bf g}):
position $x$ as a function of
initial position $x_{0}$. Parameters: $\st=300$, $\ku=0.1$.}
\end{figure}
\begin{figure}[t]
\hfill \includegraphics[width=6cm]{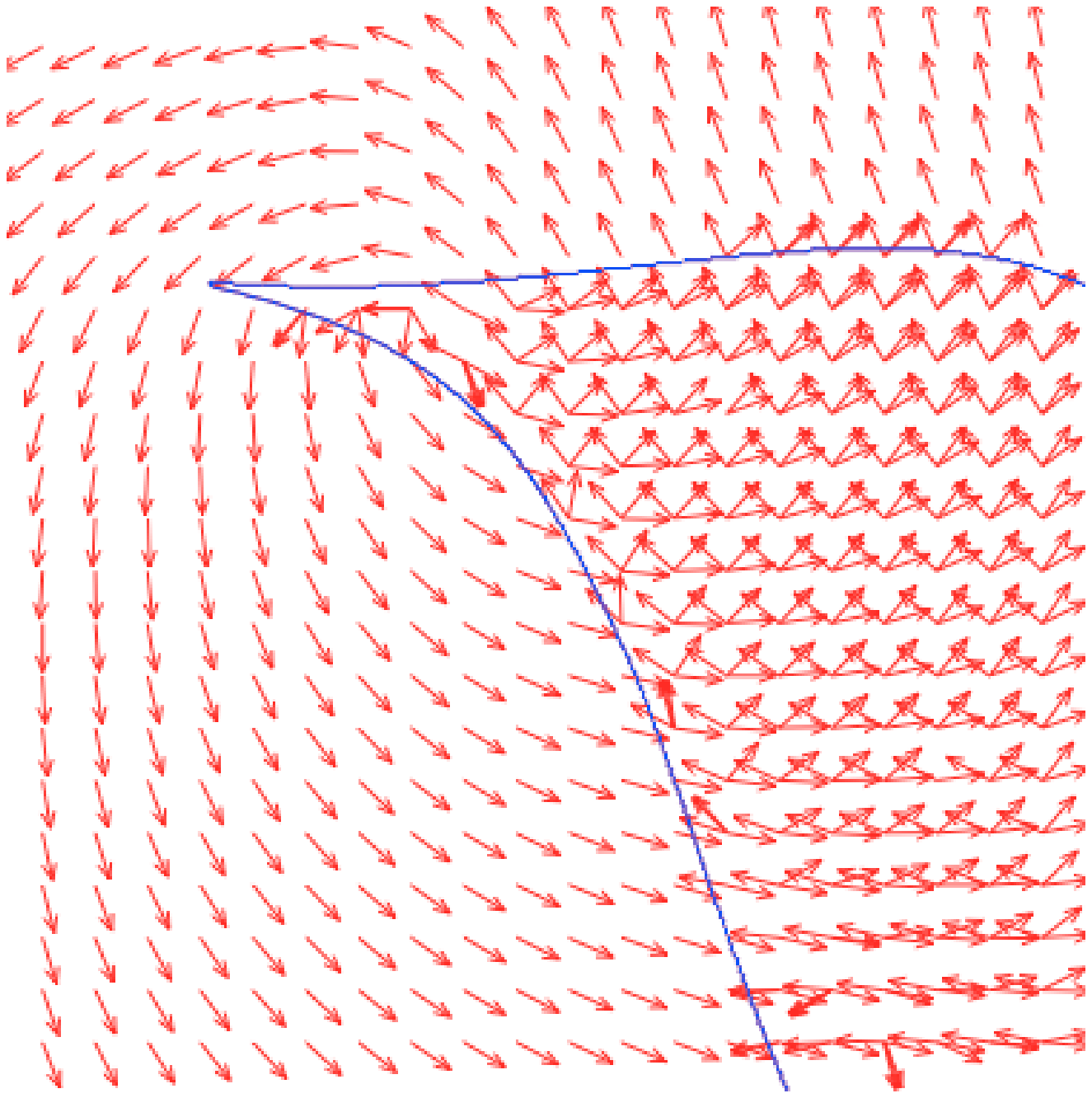}
\includegraphics[width=6cm]{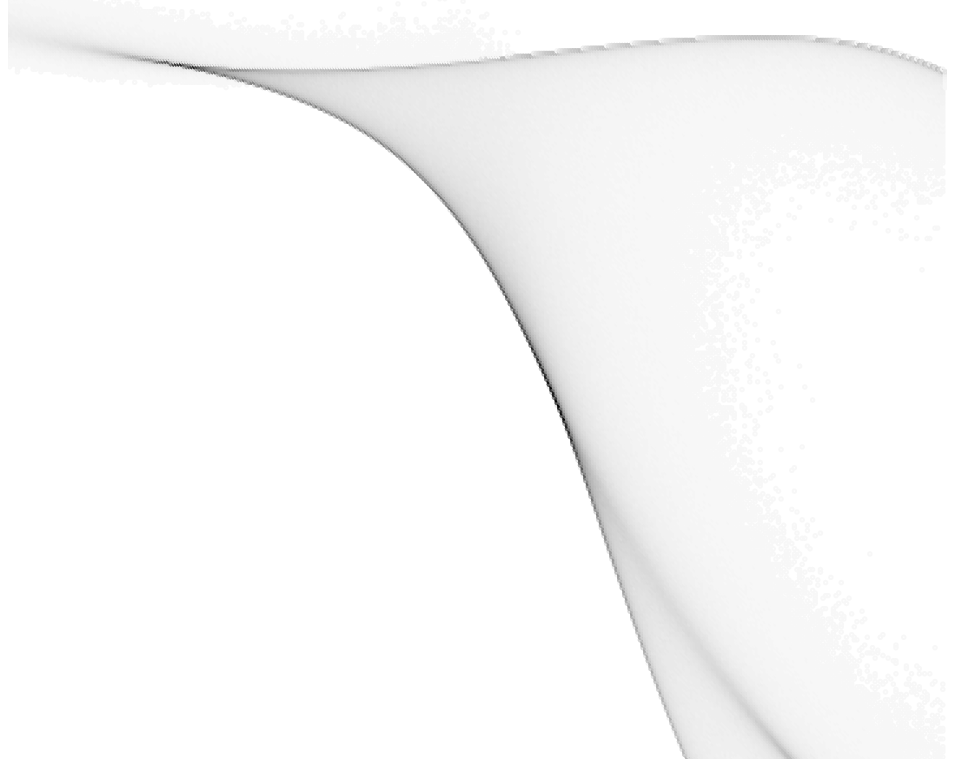}
\caption{\label{fig:2} (Left) multi-valued velocities of particles suspended in a
two-dimensional random flow with finite $\ku$ and $\st$ as described
in Section~\ref{ssec:random}. The base of each red arrow corresponds
to a particle position (taken to be on a regular grid in the $x$-$y$-plane).
The orientation of the velocity is that of the arrow. All arrows
have the same length, the magnitudes of the velocities
are not shown. The blue line delineates the position of the caustics
in the $x$-$y$-plane. The region of multi-valued velocities ends
in a cusp that is only approximately resolved. In Section \ref{sec:RUM}
it is explained how multi-valued velocities between caustics
give rise to so-called random uncorrelated motion. Parameters: $\ku=1$, $\st=10$.
(Right) particle-density in the $x$-$y$-plane, showing significantly
enhanced particle-number density in the vicinity of the caustic line.
Same parameters as above. Black corresponds to high density, white to low density. }
\end{figure}

Arguably spatial clustering has an effect upon
the collision rate at small Stokes numbers. But there is
a second mechanism that leads to a significant enhancement
of the collision rate as the Stokes number increases:
direct numerical simulations of particles suspended in turbulent
flows \cite{Sun97,Wan00} show that relative particle velocities at
small separations increase substantially as the Stokes number is varied
beyond a threshold of order unity. In \cite{Fal02,Wil06} this behaviour 
was explained by the occurence of  singularities in the particle
dynamics, causing large relative velocities at small separations.
These singularities occur as the phase-space manifold 
folds, as illustrated in Fig.~\ref{fig:1}.
As a consequence, particle velocities at 
a given point in space become multi-valued, causing large velocity
differences between nearby particles.
The boundaries of the folding region are referred to as \lq caustics'
\cite{Wil05,Cri92}. It was shown
that the
rate of caustic formation is an activated process \cite{Wil05,Dun05,Gus12}.
This explains
the sensitive dependence of the rate of caustic formation
upon the Stokes number observed in direct numerical simulations of particles in turbulence \cite{Pum07}.

An alternative way of characterising relative velocities of inertial
particles was suggested in \cite{Fev05,Sim06}. The authors of these
papers decomposed inertial particle velocities into two
contributions: a spatially correlated, smoothly varying \lq filtered' velocity field, and a random,
spatially and temporally uncorrelated contribution,
commonly referred to as \lq random uncorrelated
motion' \cite{Ree06,Mas11}.

The aim of this paper is twofold. First we summarise
results of numerical simulations of particles suspended
in model flows (Figs.~\ref{fig:mp_WN} - \ref{fig:C1_kinetic}). Our numerical results 
for the moments of relative velocities of inertial particles
are in quantitative agreement with recent analytical results 
based on the notion of caustic formation \cite{Gus10}.
Second we demonstrate that
caustic formation not only provides an understanding of relative velocities
at small separations, caustic formation also explains spatial clustering
due to singularities in the local deformation tensor, and the existence
and properties of random uncorrelated motion.

We conclude the introduction by summarising our results in more detail.
In this paper we show that recent predictions by Wilkinson~et~al.~(2006)
and Gustavson \harvardand\ Mehlig (2011{\em a})  based on
the notion of caustic formation describe many aspects of the fluctuations
of relative velocities at small separations. We compare formulae for
the moments of relative velocities (Eqs.~(\ref{eq:mp_1d}) and
(\ref{eq:mp}) below) to new results of numerical simulations of one- and  two-dimensional models 
for inertial particles suspended in white-noise flows, and
for a three-dimensional kinematic simulation of particles suspended in an incompressible flow field
with an energy spectrum typical of the small scales of turbulence.
We find good agreement. This demonstrates that Eqs.~(\ref{eq:mp_1d}) and (\ref{eq:mp})
which were derived in the white-noise limit, are valid more generally.

Further we examine the prediction by \cite{Fev05,Sim06}
that the so-called longitudinal second-order structure function of
relative velocities tends to a finite value at vanishing separations
in the presence of random uncorrelated motion.
The analytical theory, Eqs.~(\ref{eq:mp_1d}) and (\ref{eq:mp})
below, shows that this is true for sufficiently large Stokes numbers
[the case examined numerically by \cite{Sim06}].
But at Stokes numbers smaller than a critical value, the structure
function tends to zero, despite the fact that there may still be a
substantial singular (multi-valued) contribution to relative velocities
due to the formation of caustics.

We discuss in detail that the singularities of the deformation
tensor are in fact caustic singularities, as pointed out by \cite{Wil07}.
We study the dynamics of the deformation tensor $\ma J$, and
the matrix $\ma Z$ of particle-velocity gradients. We show that as
$\det\ma J$ approaches zero, $\tr\ma Z\rightarrow-\infty$. We briefly
remark upon the statistical properties of the singularities \cite{Men11}.

In summary, we demonstrate that the notion of random uncorrelated
motion, and the occurrence of zeroes in the local deformation tensor
can both be explained in terms of caustic formation, both qualitatively
and in many ways quantitatively. Last but not least our results indicate
that the white-noise approximation successfully describes many aspects
of turbulent aerosols.

 The remainder of this paper is organised as follows. In Section~\ref{sec:model}
 we introduce the models analysed in this paper: inertial particles
 suspended in a two-dimensional incompressible random flow in the white-noise
 limit, and a kinematic simulation of inertial particle dynamics. Section~\ref{sec:caustic} summarises
 what is known about the rate of caustic formation and discusses
 consequences for the fluctuations of relative particle velocities.
We compare the analytical theory to results of numerical simulations
of the models described in Section~\ref{sec:model}. In Section~\ref{sec:RUM}
we briefly review the notion of random uncorrelated motion, and compare
the conclusions of \cite{Fev05,Sim06} to our analytical and numerical
results. In Section~\ref{sec:J} we describe the dynamics of the
local deformation tensor and its correspondence to the dynamics of the matrix of particle-velocity gradients.
Finally, Section~\ref{sec:conc} contains our conclusions.

\section{Model}

\label{sec:model} The motion of small, non-interacting spherical
particles suspended in a flow is commonly approximated by
\begin{equation}
\dot{\ve r}=\ve v\,,\hspace{1cm}\dot{\ve v}=\gamma(\ve u-\ve v)\,.\label{eq:stokes_dim}
\end{equation}
 Here $\ve r$ and $\ve v$ are the position and velocity of a particle,
$\ve u(\ve r,t)$ is the velocity field evaluated at the particle
position, $\gamma$ is the viscous damping rate, and dots denote time
derivatives. The components of the vector $\ve r$ are denoted by
$r_{j}$, $j=1,\ldots,d$ in $d$ dimensions. The components of $\ve u$
and $\ve v$ are referred to in an analogous way. Sometimes it is
more convenient to denote the components of $\ve r$ by $(x,y,z)$
instead of $(r_{1},r_{2},r_{3})$. We use the two notations interchangeably.

For Eq.~(\ref{eq:stokes_dim}) to be valid, it is assumed that the
particle Reynolds number is small, that Brownian diffusion of the
particles can be neglected, and that the particle density is much larger
than that of the fluid. We also assume that the velocity field $\ve u$
varies smoothly on small spatial and temporal scales with smallest
length-and time scales $\eta$ and $\tau$ (the Kolmogorov scales
for turbulent flows). The typical magnitude of the velocity field
is denoted by $u_{0}$.

In dimensionless units ($t=t'/\gamma$, $\ve r=\eta\ve r'$, $\ve v=\gamma\eta\ve v'$,
$\ve u=\gamma\eta\ve u'$ and dropping the primes), the equation of
motion becomes
\begin{equation}
\dot{\ve r}=\ve v\,,\hspace{1cm}\dot{\ve v}=\ve u-\ve v\,.\label{eq:stokes}
\end{equation}
The Stokes number does not appear explicitly in this equation,
but the fluctuations of the dimensionless velocity $\ve u$ depend
upon $\st$ (see Eq.~(\ref{eq:corr}) below). In addition to the Stokes number,
the dynamics is characterised by a second dimensionless
number, the \lq Kubo number\rq~$\ku\equiv u_{0}\tau/\eta$. We
note that turbulent flows have $\ku\sim1$. In the remainder of this
paper we frequently refer to these two dimensionless numbers.
For a discussion of further dimensionless parameters see \cite{Wil07}.
The numerical results shown in the following were obtained for two
different models.
These models are introduced in the following two subsections.

\subsection{Random-flow model}
\label{ssec:random} Following \cite{Wil03,Wil05,Dun05,Wil07} we
approximate the incompressible velocity field $\ve u(\ve r,t)$ in Eq.~(\ref{eq:stokes})
by a Gaussian random function that varies smoothly in
space and time. We discuss results for one- and two-dimensional
versions of the random-flow model. The one-dimensional case is most
easily analysed, the two-dimensional incompressible case is important
(since one-dimensional flows are special, they are always compressible
which give rise to a path-coalescence transition \cite{Wil03}). A two-dimensional
incompressible velocity field can be written in terms of a stream
function $\psi(\ve r,t)$:
\begin{equation}
\ve u(\ve r,t)=\ve\nabla\wedge\psi(\ve r,t){\bf e}_{3}\,.
\end{equation}
 Here ${\bf e}_{3}$ is the unit vector $\perp$ to the $x$-$y$-plane.
We assume that  $\psi(\ve r,t)$  is a Gaussian random function
with $\langle\psi\rangle=0$ and correlation function
\begin{equation}
\label{eq:corr}
\langle\psi(\ve r,t)\psi(\ve0,0)\rangle=
{1\over 2} \ku^2 \st^2
 \exp\left[-|\ve r|^{2}/2-\st\,|t|\right]\,.
\end{equation}
in dimensionless variables.

In this paper we also refer to results of a one-dimensional
random-flow model. This is defined in an analogous fashion
in terms of a Gaussian random flow velocity $u(x,t)$ with
zero mean and correlation function
\begin{equation}
\langle u(x,t)u(0,0)\rangle=
\ku^2 \st^2 \exp\left[-x^{2}/2-\st\,|t|\right]\,.
\end{equation}
We note the one-dimensional flow is compressible.
The numerical data shown in Figs.~\ref{fig:1} and \ref{fig:2} are
obtained by computer simulations of the models described above.

We simplify the model by linearising Eq.~(\ref{eq:stokes}).
This yields the following equation for the dynamics of a small separation
$\ve R=\ve r_{1}-\ve r_{2}$ and velocity difference $\ve V=\ve v_{1}-\ve v_{2}$
between two particles:
\begin{equation}
\dot{\ve R}=\ve V\,,\hspace{1cm}\dot{\ve V}=\ma A\ve R-\ve V\,.\label{eq:stokes_sep}
\end{equation}
 Here $\ma A$ is the matrix of fluid velocity gradients, with elements
$A_{ij}={\partial u_{i}}/{\partial r_{j}}$.

To simplify further, we take the white-noise limit of this model.
This limit corresponds to
\begin{eqnarray}
 &  & \ku\to0\quad\mbox{and}\quad\st\to\infty\quad
\mbox{such that}\quad\epsilon^{2}\equiv c_{d}\,\ku^{2}\st={\rm const.}
\end{eqnarray}
Here $\epsilon$ is a dimensionless measure of the particle inertia
introduced by \cite{Meh04} [see also \cite{Wil07}].
We take $c_{1}=1$ for one-dimensional flows
[this is consistent
with the convention used in \cite{Gus10}].
For incompressible two-dimensional flows we take $c_{2}=1/2$, as in \cite{Gus11}.
In the white-noise limit,
the instantaneous value of the velocity gradient $\ma A$ in (\ref{eq:stokes_sep})
becomes independent of the particle position. In two spatial dimensions,
we denote the independent random increments of the elements $A_{11}$, $A_{12}$ and $A_{21}$ of
$\ma A$ in a small time step $\delta t$
by $\delta a_{1},\delta a_{2}$, and $\delta a_{3}$. Note that $A_{22}=-A_{11}$ since the flow is incompressible.
One finds:
\begin{eqnarray}
\langle \delta a_{k}\rangle & = & 0\\
\langle \delta a_{k}\delta a_{l}\rangle & = & 2\epsilon^{2}\delta t\,\left(\begin{array}{ccc}
1 & 0 & 0\\
0 & 3 & -1\\
0 & -1 & 3
\end{array}\right)\,.
\end{eqnarray}
The results shown
in Figs.~\ref{fig:mp_WN}, \ref{fig:d2}, and \ref{fig:C1_WN} are
obtained by computer simulations of this model, approximating
the time-dependence of $\ma A(\ve r(t),t)$ as a white-noise signal.

\subsection{Kinematic simulation}
\label{ssec:kinetic}
As an alternative to the single-scale white-noise model introduced
in the previous subsection, we simulate a turbulent
incompressible velocity field in a three-dimensional periodic box
by a large number of Fourier modes varying randomly in space and time.
The modes are chosen in such a way that the associated energy spectrum
approximates a prescribed form, namely that originally used by \cite{Kra70}.
The model is identical to that used by \cite{Ijz10,Men11}.
For convenience we briefly summarise its relevant features below.
For details, we refer the reader to \cite{Ijz10,Men11}.

In dimensionless form, the incompressible velocity field $\ve u(\ve r,t)$ is represented as a Fourier
series of $N$ modes ($N=200$ in our simulations):
\begin{eqnarray}
\ve u(\ve r,t)&=&\sum_{n=1}^{N}\Big[\frac{\ve a^{(n)}\wedge \ve k^{(n)}}{|\ve k^{(n)}|}\cos\Big(\ve k^{(n)}\cdot\ve r
+\omega^{(n)}t\Big)\nonumber\\ &&
+\frac{\ve b^{(n)}\wedge \ve k^{(n)}}{|\ve k^{(n)}|}\sin\Big(\ve k^{(n)}\cdot\ve r+\omega^{(n)}t\Big)\Big],\label{eq:flow_field_KS_Fourier_modes}
\end{eqnarray}
 with random coefficients $\ve a^{(n)}$ and $\ve b^{(n)}$,
random wave numbers $\ve k^{(n)}$, and random frequencies
$\omega^{(n)}$.
In order to guarantee the periodicity
of the flow in a cube of dimensions $L\times L\times L$, the
allowed wave number components $k_{i}^{(n)}(i=1,2,3)$ are
\begin{equation}
k_{i}^{(n)}=\frac{2\pi m_{i}^{(n)}}{L}
\label{eq:select_k_1_2_3_periodic}
\end{equation}
with $ m_{i}^{(n)}=0,\pm 1,\pm 2,\ldots$.
We take $L=10\,L_{\rm int}$, where $L_{\rm int}=\sqrt{2\pi}$ is the integral length scale of the flow.
The integer numbers $m_{i}^{(n)}$ are chosen randomly
in such a way that the lengths $k^{(n)}=\sqrt{\ve k^{(n)}\cdot\ve k^{(n)}}$
are approximately equal to the ideal wave number $k_{\rm id}^{(n)}$.
The latter is determined by the energy spectrum as follows:
\begin{equation}
\int_{0}^{k_{\rm id}^{(n)}}\!\!{\rm d}k\,E(k)=\frac{3}{2}\frac{(n-1/2)}{N}\,.
\end{equation}
As mentioned above, the energy spectrum $E(k)$ is taken to be \cite{Kra70}:
\begin{equation}
\label{eq:kra}
E(k)=\frac{32\,k^4}{\sqrt{2\pi}}\exp\left(-2k^{2}\right)\,.
\end{equation}
This spectrum is representative for low-Reynolds-number turbulence \cite{Spe97}.
The maximum of $E(k)$ is located at $k=1$ and the total kinetic
energy $\int_{0}^{\infty}E(k)\mbox{d}k=3/2$. This corresponds to  $3u_{0}^{2}/2$ in dimensional form. The use of the
Kraichnan energy spectrum results
in a relatively small separation of scales; in our simulations, the
smallest wave number $k^{(1)}\simeq0.25$ and the largest wave number
$k^{(N)}\simeq2.14$.
The frequencies $\omega^{(n)}$ are chosen randomly from a
Gaussian distribution with zero mean and a variance
proportional to $k^{(n)}$. This implies that the Kubo
number is of order unity.
Following \cite{Spe97}, we take the variance to be $0.4\,k^{(n)}$.
Finally, the coefficients $\ve a^{(n)}$ and $\ve b^{(n)}$
are determined by choosing a random direction in Cartesian space,
and by picking a length randomly from a Gaussian distribution with
zero mean and a variance $9/(2N)$. By doing so, the mean kinetic energy
at a given position in space
\begin{eqnarray}
 \bar{E}_{kin}(\ve r)&=&\frac{1}{T}\lim_{T\rightarrow\infty} \frac{1}{2}\int_{0}^{T}\!\!{\rm d}t\,|\ve u(\ve r,t)|^{2}\\
 &=&\sum_{n=1}^{N}\frac{1}{4|\ve k^{(n)}|^{2}}\biggl[|\ve a^{(n)}\wedge\ve k^{(n)}|^{2}+|\ve b^{(n)}\wedge\ve k^{(n)}|^{2}\biggr],
\label{eq:total_kinetic_energy_KS}\nonumber
\end{eqnarray}
is approximately equal to $3/2$ for all values of $\ve r$.

\section{Caustics}

\label{sec:caustic}

\subsection{One spatial dimension}

As illustrated in Fig.~\ref{fig:1}, caustics form when the phase-space
manifold folds over. In one spatial dimension this happens when the
slope of the manifold becomes infinite, that is when $z={\partial v}/{\partial x}\rightarrow-\infty$.
The rate at which this occurs is determined by the equation of motion
for $z$ \cite{Wil03}:
\begin{equation}
\dot{z}=A-z-z^{2}\,.\label{eq:zEq_1d}
\end{equation}
 Here $A=\partial u/\partial x$ represents the random driving by
the fluid-velocity gradients. In the case of independent particles
(which we consider here), $z$ goes through infinity in a symmetrical
fashion. At large values of $|z|$, the random driving can be neglected,
so that $\dot{z}\approx-z-z^{2}$. The corresponding deterministic
probability distribution of $z$ reads $\rho(z)=C/[z(1+z)]$, and
is valid in the tails of $z$.

In the white-noise limit, Eq.~(\ref{eq:zEq_1d}) is equivalent to
a Fokker-Planck equation for the distribution of $z$. In \cite{Wil03}
this equation was solved in one spatial dimension. The resulting rate
of caustic formation [called \lq rate of crossing caustics' by
Wilkinson \harvardand\ Mehlig (2003)] can be written as \cite{Gus12}:
\begin{equation}
\frac{J_{\rm caustic}}{\gamma}=\frac{1}{2\pi}{\mathcal{I}m}\Big[\frac{{\rm Ai}'(y)}{\sqrt{y}{\rm Ai}(y)}\Big]\Bigg|_{y=(-1/(8\,\epsilon^{2}))^{2/3}}\,,\label{eq:caustic_rate_1d}
\end{equation}
 where $\epsilon^{2}=\ku^{2}\st$ (see Section \ref{ssec:random}).
In Eq.~(\ref{eq:caustic_rate_1d}), $\rm Ai(y)$ is the Airy function.
In the limit of small values of $\epsilon$, this expression exhibits
the asymptotic behaviour 
\begin{equation}
\frac{J_{\rm caustic}}{\gamma}\sim\frac{1}{\sqrt{2\pi}}{\rm e}^{-1/(6\epsilon^{2})}\,.\label{eq:caustic_rate_1d_asy}
\end{equation}
 Eq.~(\ref{eq:caustic_rate_1d}) shows that 
the number of caustics increases rapidly as $\epsilon^{2}$ passes
through $1/6$ \cite{Wil05,Wil06}. This sensitive dependence
is commonly referred to as an \lq activated law', in analogy
with the sensitive temperature depdence of chemcial reaction rates
in Arrhenius' law.  Gustavsson \harvardand\ Mehlig (2012)  computed the one-dimensional rate of caustic formation
at small but finite Kubo numbers and found it to sensitively depend
on the Stokes number: in this case too, the $\st$-dependence exhibits
\lq activated form': $J_{\rm caustic}/\gamma\sim\exp[-S(\st)/\ku^{2}]$, where
$S$ is an $\st$-dependent \lq action\rq. In the white-noise limit,
$S=1/(6\st)$, consistent with Eq.~(\ref{eq:caustic_rate_1d_asy}).

As Fig.~\ref{fig:1} shows, particle-velocities become multi-valued
between two caustics in the wake of a singularity, giving rise to
large relative velocities between nearby particles. While the rate
of caustic formation is determined by the rate at which the local
quantity $z=\partial v/\partial x$ tends to $-\infty$, the distribution
of relative velocities at small particle separations is determined
by the solution of the full non-local equations
(\ref{eq:stokes_sep}) for particle separations and relative
velocities \cite{Gus10}.

A consequence of large relative velocities at small separations is
that between caustics, particles collide frequently with large relative
velocities (c.f. Fig.~\ref{fig:1}), giving rise to a large collision
rate (we note, however, that in this paper it is assumed that the
particles are independent point particles that do not actually collide).

By contrast, in the absence of caustics, particles may still approach
each other due to fluctuations of the underlying flow-velocity field.
At small separations the flow is smooth, and in this regime relative
velocities between particles are expected to tend to zero as the particles
in question approach each other.

Which one of these two mechanisms of bringing particles together makes
the dominant contribution to the collision rate depends upon the value
of $\st$ and on the particle size $a$ (separation $2a$ at the point
of contact). Relative velocities of particles thrown at each other
due to the formation of caustics are expected to make the dominant
contribution to the collision rate if $\st$ is large and/or when
the particles are sufficiently small. Particles slowly approaching
each other (\lq logarithmic diffusion') dominate otherwise.

In the white-noise limit, and in one spatial dimension, an asymptotic
approximation for the moments of relative velocities at small separations 
was derived in \cite{Gus10}: 
\begin{eqnarray}
m_{p}(X) & = & \int_{-\infty}^{\infty}{\rm d}V|V|^{p}\rho(X,V)  \sim  B_{p}|X|^{p+D_{2}-1}+C_{p}\,.\label{eq:mp_1d}
\end{eqnarray}
Here $X = x_1-x_2$ and $V = v_1-v_2$ are the separation and the relative velocity of a pair of particles, and $\rho(X,V)$
is their distribution function. It is assumed that $|X|\ll 1$ and $p> -1$. Further,
$D_{2}$ is the correlation dimension of the phase-space attractor
and $B_{p}$ and $C_{p}$ are model-dependent constants [which were
not derived by \cite{Gus10}]. The form of Eq.~(\ref{eq:mp_1d})
is consistent with the form inferred from simulations of relative-particle
dynamics in a one-dimensional Kraichnan model \cite{Cen09}. 

The second term in Eq.~(\ref{eq:mp_1d}), $C_{p}$, is due to multi-valued velocities between caustics. This contribution, in one spatial dimension, 
does not depend upon $|X|$ for small values of $|X|$. In other words: it remains
finite as $|X|\rightarrow 0$. This is a consequence of the fact that
as the manifold in Fig.~\ref{fig:1} folds over, particles initially far apart are thrown at each other
quickly. 

The first term in Eq.~(\ref{eq:mp_1d}), $B_{p}|X|^{p+D_{2}-1}$, vanishes as $|X| \rightarrow 0$. 
It constitutes the main contribution to $m_p(X)$ in the absence of caustics and is
affected by spatial clustering:
for a given value of $p$, the exponent
is smallest (and thus the contribution largest) when $D_{2}$ attains
its minimum as a function of Stokes number.

In Eq.~(\ref{eq:mp_1d}) the case $p=1$ is of particular importance, since $m_{1}(X)$ is
closely related (yet not identical) to the collision rate between
particles at small separations $X=2a$. It is expected that the coefficient
of the caustic contribution in Eq.~(\ref{eq:mp_1d}), $C_{1}$, is
proportional to the caustic formation rate $J_{\rm caustic}$ \cite{Wil06}.

\subsection{Two and three spatial dimensions}
\begin{figure}[t]
\psfrag{x}{\large $R$}
\psfrag{y}{\raisebox{2mm}{\large $m_{p}(R)$}}
\hfill \includegraphics[width=7cm]{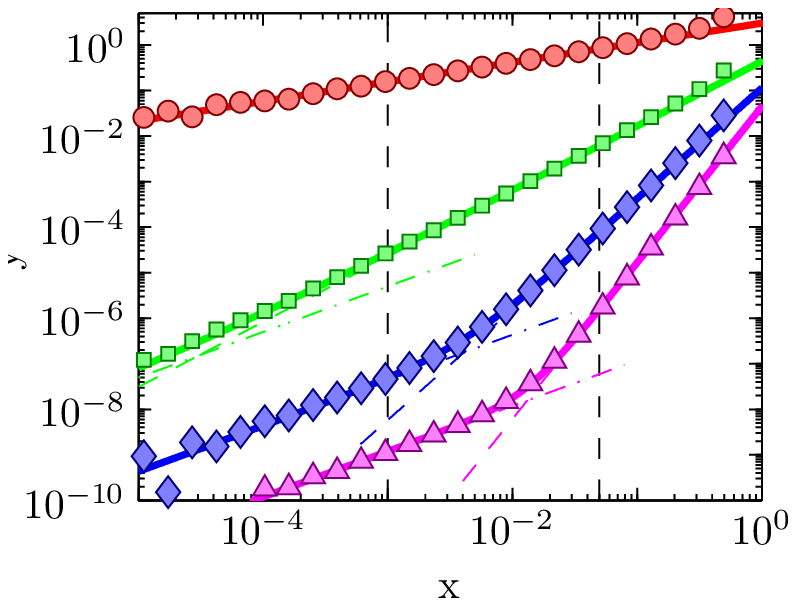}\hspace*{5mm}
\includegraphics[width=7cm]{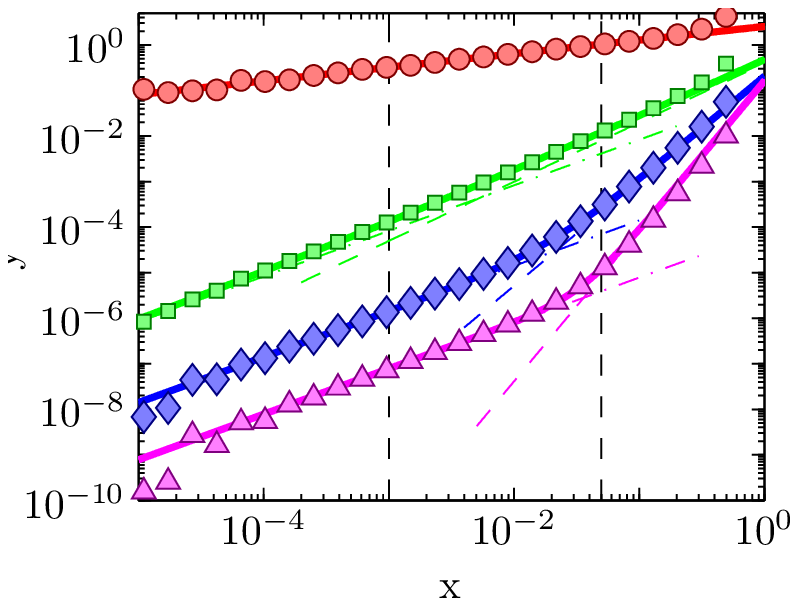}
\caption{\label{fig:mp_WN} Moments of the radial velocity $m_{p}(R)$
plotted against distance $R$ for two different values of $\epsilon$:
$\epsilon=0.03$ (left) and $\epsilon=0.06$ (right). Data from numerical
simulations of the two-dimensional white-noise model described in
Section~\ref{ssec:random} are shown as markers. The correlation
dimension $d_{2}$ and the coefficients $B_{p}$ and $C_{p}$ in the
small $R$ approximation (\ref{eq:mp}) are numerically fitted to the
data in the interval bounded by vertical black dashed lines. The resulting
moments for small $R$ (\ref{eq:mp}) are shown as solid lines. The
caustic contribution $C_{p}R^{d-1}$ (dashed dotted) and the smooth
contribution $B_{p}R^{p+d_{2}-1}$ (dashed) are also shown. Parameters:
$p=0$ (red $\circ$), $p=1$ (green $\Box$), $p=2$ (blue $\Diamond$)
and $p=3$ (magenta $\vartriangle$).}
\end{figure}

In two and three spatial dimensions the caustic rate can be found in
a way similar to the one-dimensional case \cite{Wil07,Gus11}:
the matrix $\ma Z$ with elements $Z_{ij}={\partial v_{i}}/{\partial r_{j}}$
obeys the equation:
\begin{equation}
\dot{\ma Z}=\ma A-\ma Z-\ma Z^{2}\,.\label{eq:zEq}
\end{equation}
Here $\ma A$ is the matrix of fluid-velocity gradients introduced
in Section \ref{sec:model}, with elements $A_{ij} = \partial u_i/\partial r_j$.
 In analogy with the one-dimensional case, $\mbox{tr}(\ma Z)\rightarrow-\infty$
as caustics are formed. In the white-noise limit, we expect that the
rate of caustic formation is again given by (\ref{eq:caustic_rate_1d_asy}).
In  \cite{Wil05,Dun05,Wil07}
numerical factors in Eq.~(\ref{eq:caustic_rate_1d_asy}) slightly different from $1/6$ were quoted
in two and three spatial dimensions. More recent numerical results (not
shown) show that the asymptote (\ref{eq:caustic_rate_1d_asy})
is approached very slowly as $\epsilon$ becomes small. Our best estimates
at the smallest values of $\epsilon$ indicate that the factor in
the argument of the exponential in (\ref{eq:caustic_rate_1d_asy}) is asymptotically
the same (equal to $1/6$) in one, two, and three spatial dimensions.

Moments of relative velocities in two and three spatial dimensions
obey laws analogous to (\ref{eq:mp_1d}). At small
separations ($R \ll1 $) Gustavsson~\harvardand~Mehlig~(2011{\em a}) found
\begin{eqnarray}
m_{p}(R) & = & \int_{-\infty}^{\infty}{\rm d}v_{R}|v_{R}|^{p}\rho(R,v_{R}) \sim  B_{p}R^{p+d_{2}-1}+C_{p}R^{d-1}\,.\label{eq:mp}
\end{eqnarray}
Here $R=|\ve R|$ and $v_{R}\equiv\ve V\cdot\hat{{\bf e}}_R$ is the radial projection of the relative velocity between two particles at separation
$\ve R$.
Further, $d_2$ is the spatial correlation dimension, it is assumed that the Stokes number is small enough so that $d_2 \leq d$.
As in the one-dimensional result,
Eq.~(\ref{eq:mp_1d}), there are two contributions to the moments
of relative velocities
[compare the parameterisation of the $\st$-dependence of the collision rate
suggested by Wilkinson et al. (2006)]. 

The second term in Eq.~(\ref{eq:mp})
is due to multi-valued velocities between caustics.
But note that in two and three spatial dimensions, not all particle pairs thrown
together give rise to close approaches. The reason is that in addition to
having one relative coordinate pass zero at finite relative velocity
(so that a caustic occurs), the other coordinates must be small, i.e. only
particles heading sufficiently towards each other as the caustic occurs
end up at small enough separations to contribute to the small $R$ velocity moments. This explains the geometrical factor $R^{d-1}$
in $(\ref{eq:mp})$. It is absent in one spatial dimension, $d=1$.

Fig.~\ref{fig:mp_WN} shows comparisons of Eq.~(\ref{eq:mp}) with
results of numerical simulations of the random-flow model described
in Section~\ref{ssec:random}. The parameter $d_{2}$ in Eq.~(\ref{eq:mp})
is determined as follows. Setting $p=0$ in (\ref{eq:mp}) and taking
the limit $R\to0$ defines the spatial correlation dimension $d_{2}$.
The latter is found numerically by fitting $m_{0}(R)$ to the power
law $R^{d_{2}-1}$.

\begin{figure}[t]
\psfrag{x}{\large $\epsilon^{2}$}
\psfrag{y}{\raisebox{0.2cm}{\large $d_{2}$}}
\hfill \includegraphics[width=7.0cm]{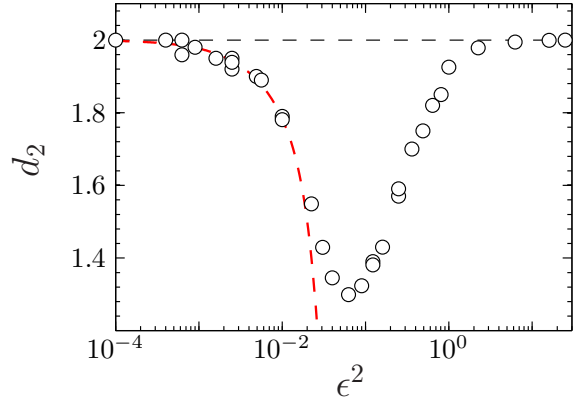}
\caption{\label{fig:d2} Spatial correlation dimension $d_{2}$ ($\circ$)
as a function of
$\epsilon^{2}$ for the model described in Section~\ref{ssec:random}.
The dashed red line shows the small-$\epsilon$ theory discussed in 
(Wilkinson et al. 2010, Bec et al. 2008).}
\end{figure}
\begin{figure}[t]
\psfrag{y}{\raisebox{0.2cm}{\large $C_{1}$}}
\psfrag{z}{\raisebox{0.6cm}{$J_{\rm caustic}/\gamma$}}
\psfrag{x}{\large $\epsilon^{-2}$}
\hfill \includegraphics[width=7.5cm]{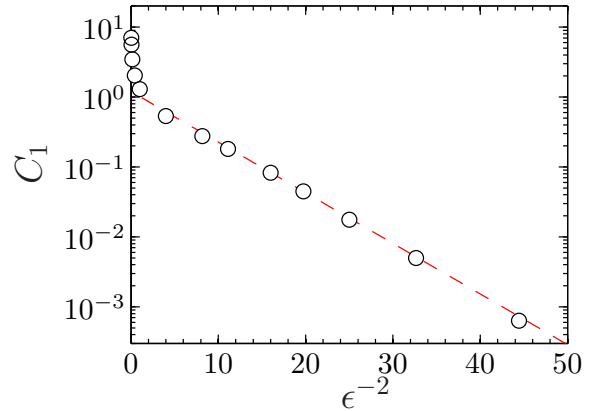}
\caption{\label{fig:C1_WN} Amplitude $C_{1}$ from numerical
fits to $m_{1}$ in (\ref{eq:mp}) (symbols) as a function of $\epsilon^{-2}$.
Numerical simulations of the model described in Section~\ref{ssec:random}.
The asymptotic $\st$-dependence of the rate of caustic formation, (\ref{eq:caustic_rate_1d_asy}),
is shown as a dashed line.}
\end{figure}

We now describe how the fits in Fig.~\ref{fig:mp_WN} were obtained.
The parameter $d_{2}$ was taken from Fig.~\ref{fig:d2}. The coefficients
$B_{p}$ and $C_{p}$ in Eq.~(\ref{eq:mp}) were fitted to the numerical
results for different parameter values. The fitting region (the range
of $R$ over which Eq.~(\ref{eq:mp}) is fitted) lies between the
dashed lines in Fig.~\ref{fig:mp_WN}. We observe good agreement
between the numerical results and fits to Eq.~(\ref{eq:mp}). In
particular, the results clearly show that the moments $m_{p}$ scale
as $R^{d-1}$ for small values of $R$, independently of $p$.
Fig.~\ref{fig:C1_WN}
shows the coefficient $C_{1}$ of the caustic contribution obtained
in this way as a function of $\epsilon^{-2}$. Since this contribution
requires the formation of caustics, we expect $C_{1}$ to exhibit
an $\epsilon$-dependence of the form (\ref{eq:caustic_rate_1d_asy}).
Fig.~\ref{fig:C1_WN} shows that this is indeed the case.

\begin{figure}[t]
 \psfrag{x}{\large $R$}
\psfrag{y}{\raisebox{2mm}{\large $m_{p}(R)$}}
\hfill \includegraphics[width=7cm]{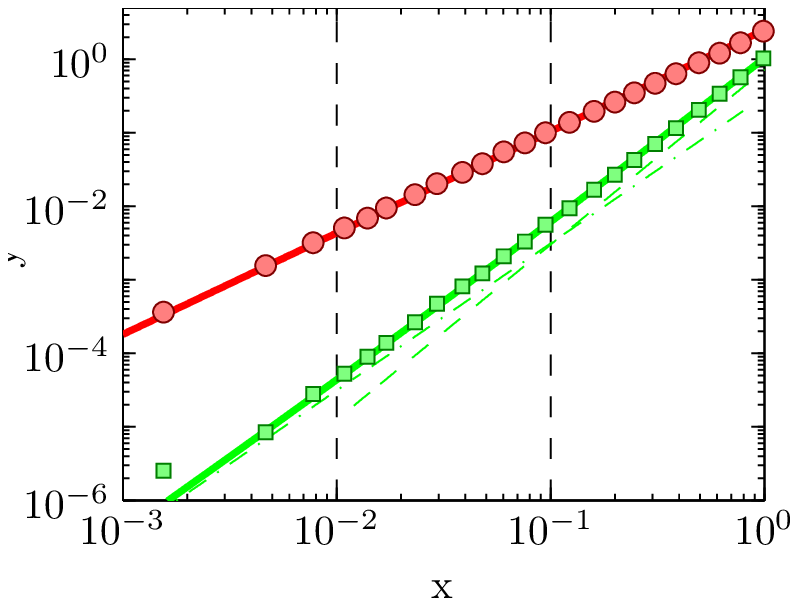}\hspace*{5mm}
 \includegraphics[width=7cm]{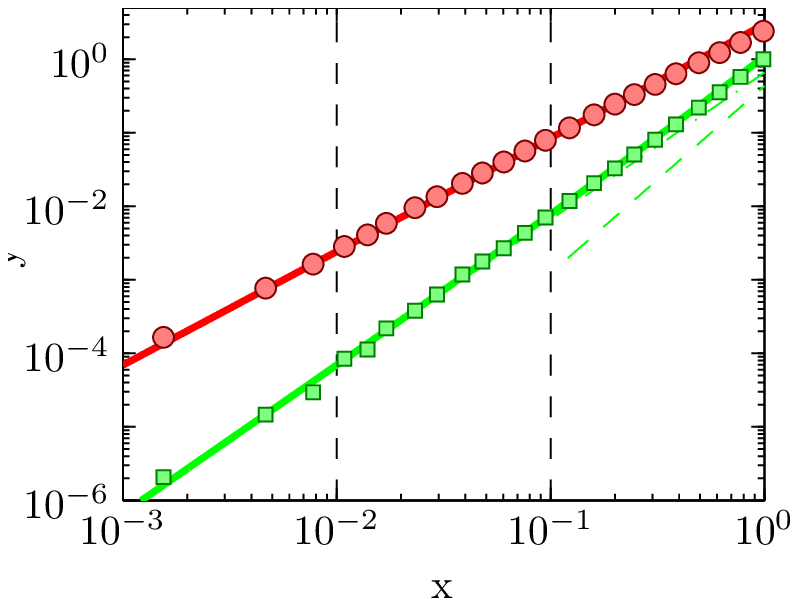}
\caption{\label{fig:mp_kinetic} Same as Fig.~\ref{fig:mp_WN}, for
the model described in Section~\ref{ssec:kinetic} with $\st=0.4$ (left) and $\st=0.7$ (right), and $p=0,1$.  }
\end{figure}
\begin{figure}[t]
\psfrag{x}{\large $\st$}
\psfrag{y}{\raisebox{0.2cm}{\large $d_{2}$}}
\hfill \includegraphics[width=7.0cm]{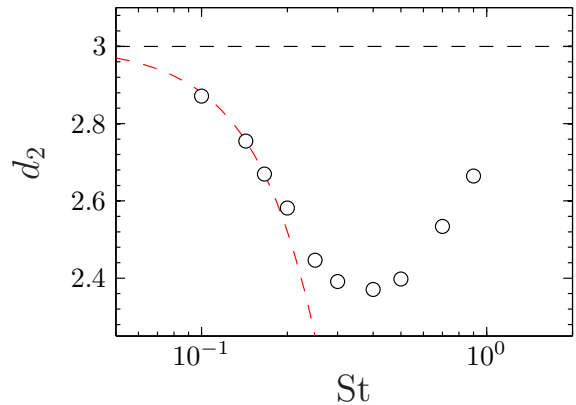}
\caption{\label{fig:d2_kinetic} Numerical results for the spatial
correlation dimension $d_{2}$ as a function of the Stokes number
for the model described in Section~\ref{ssec:kinetic}.
The dashed line shows a fit of the form $d_2 = 3-12\, \st^2$.}
\end{figure}

Fig.~\ref{fig:mp_kinetic} shows results for $m_{0}(R)$ and $m_{1}(R)$
obtained by kinematic simulations of the random-flow model described in Section~\ref{ssec:kinetic}
for two values of the Stokes number, $\st=0.4$ and $\st=0.7$.
As expected, $m_{0}(R)$ is of power-law form, reflecting
spatial clustering. The corresponding correlation dimensions are shown,
as a function of $\st$, in Fig.~\ref{fig:d2_kinetic}. The correlation
dimension exhibits the expected minimum (here at $\st\approx 0.4$). Corresponding
results for direct numerical simulations of particles in turbulent
flows have been obtained by a number of authors \cite{Wil10,Bec10,Chun}.

The green squares in Fig.~\ref{fig:mp_kinetic} correspond to numerical
results for $m_{1}(R)$ as a function of $R$.
Consider first the left panel ($\st=0.4$). At small separations
$R$ we expect that $m_{1}(R)$ should scale as $R^{d-1}=R^{2}$,
while it should scale as $R^{d_{2}}\approx R^{2.4}$ at large values
of $R$. Despite the fact that the two powers are rather similar,
the two scalings can be distinguished in Fig.~\ref{fig:mp_kinetic}.
In the right panel ($\st=0.7$), the caustic contribution $R^{d-1}$
dominates.

\begin{figure}[t]
\psfrag{x}{\large $\st^{-1}$}
\psfrag{y}{\raisebox{0.2cm}{\large $C_{1}$}}
\psfrag{z}[cc][][1][180]{\raisebox{0.6cm}{$J_{\rm caustic}/\gamma$}}
\hfill \includegraphics[width=7.5cm]{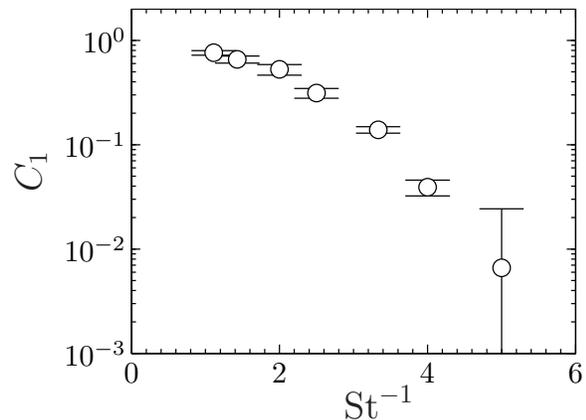}
\caption{\label{fig:C1_kinetic} Amplitude $C_{1}$ from
numerical fits to $m_{1}$ in (\ref{eq:mp}) as a function of $1/\st$
according to numerical simulations of the model described in Subsec.~\ref{ssec:kinetic}.}
\end{figure}

Given the data available from the kinematic simulations, it is more difficult to
reliably determine the $\st$-dependence of $C_{1}$ by fitting (solid
green line in Fig.~\ref{fig:mp_kinetic}). Our best estimates are
shown in Fig.~\ref{fig:C1_kinetic}.
The fits and the corresponding error bars were obtained
by a non-linear least-squared fit using MATLAB 2011.
We find that $C_{1}$ depends very sensitively
on $\st$, as expected because the formation of caustics is an activated
process. We expect \cite{Gus10,Gus12} that the $\st$-dependence of $C_{1}$ follows
the law $J_{\rm caustic}/\gamma\sim\exp[-S(\st)/\ku^{2}]$. However, the range of
Stokes numbers for which $C_{1}$ can reliably be estimated is too
small to determine the form of the function $S(\st)$.

Fig.~\ref{fig:C1_kinetic} demonstrates that the magnitude of relative
velocities at small separations depends very sensitively on the Stokes
number. We argue that this is a consequence of the sensitive $\st$-dependence
of the rate of caustic formation. This explains the sensitive
dependence on the Stokes number of collision velocities and collision rates
of particles suspended in turbulent flows
\cite{Sun97,Wan00,Zai}.

\section{Random uncorrelated motion}
\label{sec:RUM}
Singularities in the inertial-particle  dynamics
(corresponding to the formation of caustics) give rise to multi-valued
particle velocities at locations in space bounded by caustics: any identical
particles that are very close may move at substantially
different velocities. Figs.~\ref{fig:1} and \ref{fig:2} illustrate
this fact in one and two spatial dimensions. This implies in particular
that the relative motion of inertial particles cannot be captured
in terms of a \lq hydrodynamic' approximation describing the particle
velocities in terms of a smooth velocity field.
In particular Fevrier~et~al.~(2005) infer
from their DNS calculations of inertial-particle motion in a homogeneous
isotropic and stationary turbulent flow field that the velocity 
of a particle 
at a position $\ve r(t)$ at time $t$ in
a single realisation of the carrier flow field $\ve u(\ve r,t)$ is
given by the sum of two components
\begin{equation}
\ve v\big(\ve r(t)\!=\!\ve r,t\big)=\overline{\ve v}(\ve r,t)+\delta\ve v\big(\ve r(t)\!=\!\ve r,t\big)\,.\label{eq:rum}
\end{equation}
Here $\overline{\ve v}(\ve r,t)$ is a smoothly varying filtered
velocity field which Fevrier~et~al.~(2005) refer to as the \lq mesoscopic
Eulerian particle velocity field'. Values for the smooth component
in any realisation are found by dividing the spatial domain into cells
and calculating the average velocity associated with the number of
particles in each individual cell (the number of particles in each
cell being sufficiently large to form a statistically stationary average).
The residual component $\delta\ve v$ is termed the 'quasi Brownian velocity distribution component'  by
Fevrier~et~al.~(2005). It is now commonly referred to as \lq random uncorrelated motion', or RUM for short
(Reeks et al. 2006, Masi et al. 2011). This residual RUM part
is assumed to be uncorrelated with the smooth part and with itself at infinitesimally
small separations in space and time. 

The existence of multi-valued velocities between caustics is
consistent with a singular contribution to the particle velocities,
of the form of Eq.~(\ref{eq:rum}). We infer that the extent of random
uncorrelated motion [its relative contribution compared to the smooth
part in Eq.~(\ref{eq:rum})] must depend sensitively on the value
of $\st$, since the rate of caustic formation exhibits this sensitive
dependence on the Stokes number.

Let us consider the implications of Eq.~(\ref{eq:rum}) and the accompanying assumptions for
the second moment of the relative radial velocity between two particles, $v_R = (\ve v_1 - \ve v_2)\cdot \hat{\bf e}_R$:
\begin{eqnarray}
\langle v_R^2 \rangle &=& \langle [(\overline{\ve v}_1+\delta \ve v_1-\overline{\ve v}_2-\delta \ve v_2)\cdot \hat{\bf e}_R]^2\rangle\nonumber\\
&=& \langle [(\overline{\ve v}_1 -\overline{\ve v}_1)\cdot \hat{\bf e}_R]^2\rangle + \langle [\delta \ve v_1\cdot \hat{\bf e}_R]^2\rangle + 
\langle [\delta \ve v_2\cdot \hat{\bf e}_R]^2\rangle\,.
\label{eq:dec}
\end{eqnarray}
This result is of the same form as Eq.~(\ref{eq:mp}). The two right-most terms in (\ref{eq:dec}) correspond to
the caustic contribution in (\ref{eq:mp}).  In other words, Eq.~(\ref{eq:mp}) provides a quantitative
prediction for the contribution of random uncorrelated motion to the moments
of relative radial velocities. Consider for example the form of the so-called \lq longitudinal structure functions' for relative
velocities of the suspended particles.
Simonin~et~al.~(2006) argue that the second-order structure function remains finite as the spatial separation $R$ between particle
velocities tends to zero.
In the notation of the previous
section, the second-order structure function is given by
\begin{equation}
s^{(2)}(R)=\frac{m_{2}(R)}{m_{0}(R)}\,.
\end{equation}
The limiting behaviour of $s^{(2)}(R)$ can be deduced from Eq.~(\ref{eq:mp}).
From this equation we see that $m_{0}(R)\sim R^{\mbox{\scriptsize min}\{d_{2},d\}-1}$.
The correlation dimension $d_2$ saturates to $d$ at a critical Stokes number, $\st_{{\rm c}}$ (c.f. Fig.~\ref{fig:d2} where $d_2=d$ for $\epsilon^2>\epsilon_{\rm c}^2\approx 1$).
For $\st>\st_{{\rm c}}$ the suspended particles
are uniformly distributed in space [see also \cite{Bec10,Collins12}].
Let us consider this case. As $R\rightarrow0$, the caustic contribution
$C_2 R^{d-1}$ to $m_{2}(R)$ dominates in Eq.~(\ref{eq:mp}). This implies that
\begin{equation}
\label{eq:s2const}
s^{(2)}(R)\rightarrow\mbox{const.}\quad\mbox{as \ensuremath{R\rightarrow0}}\,,
\end{equation}
as argued in \cite{Sim06}.  For $\st<\st_{{\rm c}}$, by contrast, we find
\begin{equation}
s^{(2)}(R) \rightarrow 0 \quad \mbox{as \ensuremath{R\rightarrow0}}\,.
\end{equation}
More precisely, $s^{(2)}(R)$ tends to zero as $g^{-1}(R)$ when $R\rightarrow0$ 
(the pair correlation function $g(R)$ is given by $g(R)=m_{0}(R)/R^{d-1}$).

We emphasise that the behaviour (\ref{eq:s2const}) of the structure function in two and three spatial dimensions
must be distinguished from the fact that the moments $m_p(X)$ of relative velocities in one spatial
dimension always approach a positive constant as $|X|\rightarrow 0$ when $\st > 0$. Indeed,
we have shown that $s^{(2)}(R)$ may approach zero as $R\rightarrow 0$,
yet multi-valued particle velocities do still give rise to a substantial
singular contribution to the moments of relative velocities, as a consequence of singularities
giving rise to caustics.

Let us compare these findings to the results shown in Fig.~3(a) in Simonin {\em et al.} (2006).
The data shown in this figure (except perhaps the data set labeled \lq 1') imply that the structure function approaches a positive
constant as $R\rightarrow 0$. We conclude that the data sets shown (possibly with the exception of \lq 1')
correspond to Stokes numbers larger than $\st_{\rm c}$. It should be noted that  Simonin {\em et al.} (2006)
define their  Stokes number $\st_L$ in terms of the integral time scale of the turbulent flow. Here and
in a large part of the literature on inertial particles in turbulent flows the Stokes number $\st$
is defined in terms of the Kolmogorov time $\tau$. Since usually $\st \gg \st_L$ it is plausible
that most data sets in Fig.~3(a) in Simonin {\em et al.} (2006) correspond to $\st > \st_{\rm c}$.

We conclude by noting that it has been shown (see \cite{Meh05,Wil07} and references cited therein)
that the maximal Lyapunov exponent describing the dynamics of inertial
particles suspended in incompressible flows is positive. This implies
that the inertial particle dynamics is chaotic. In the limit of very
large Stokes numbers, inertial particle dynamics is thus similar to the
random motion of molecules in a gas [gas-kinetic limit, see \cite{Abr75}].
This justifies the view that there is a random uncorrelated component
to the inertial particle dynamics. It is a consequence of the formation
of caustics.

\section{Singularities in particle concentration}
\label{sec:J} 
\begin{figure}[t]
\psfrag{d}{\large $\delta$}
\hfill \includegraphics[width=7cm]{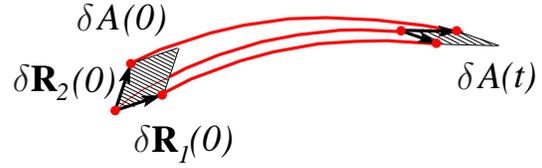}
\caption{\label{fig:element} Illustrates how an infinitesimal area element
$\delta A(t)$
in two spatial dimensions
spanned by the separation vectors $\delta\ve R_{1}$ and $\delta\ve R_{2}$
between three initially close particles is transported along the particle
trajectories.}
\end{figure}

Changes to the local concentration of inertial particles
suspended in mixing flows can be described by the deformation tensor
$\ma J$ with elements $J_{ij}={\partial r_{i}}/{\partial r_{j}(0)}$
evaluated along a particle trajectory $\ve r(t)$ with initial position
$\ve r(0)$. The matrix $\ma J$ describes the relative motion of
infinitesimally close particles. In particular, the volume spanned
by the separation vectors between $d+1$ infinitesimally close particles
in $d$ spatial dimensions is given by $\delta{\cal V}=|J|\delta{\cal V}_{0}$,
where $J\equiv\det(\ma J)$ and $\delta{\cal V}_{0}$ is the initial
volume, see Fig. \ref{fig:element}.

Nothing prevents $J$ from occasionally changing sign. This implies
that the volume $\delta{\cal V}$ may shrink to zero, giving rise to a singularity
in the local particle concentration $\propto\delta{\cal V}^{-1}$
\cite{Wil05,Wil07,Ijz10}.
The singularities influence the tails
of the distribution of  local
particle concentration, making particle clustering highly non-Gaussian and intermittent
\cite{Men11}.
The zeroes of $J$ correspond to the formation of caustics \cite{Wil07}. 
This fact is illustrated in Fig.~\ref{fig:1}: as $J\rightarrow0$ we see that $z\rightarrow-\infty$. In the following
we discuss the dynamics of $z$ and $J$ in one spatial dimension,
and then the dynamics of $\ma Z$ and $\ma J$ in two and three spatial
dimensions.

Singularities in the local particle density due to caustics occur
also in a collisionless medium of weakly interacting particles. 
As a model for the early structure of the universe, 
the corresponding linear equation of motion $\ve r(t) = \ve r_0 + t \,\ve v(\ve r_0)$
has been analysed by Zeldovich and collaborators. For a review
and a discussion of the connection between this problem and Burgers' equation
see \cite{Sha89}.

\subsection{One spatial dimension}

In one spatial dimension we analyse the joint dynamics of $z={\partial v}/{\partial x}$
and $J={\partial x}/{\partial x_{0}}$, where $x_{0}=x(0)$ is the
initial particle position. Noting that $\dot{J}={\partial v}/{\partial x_{0}}$
we see that $z=\dot{J}/J$. The dynamics of $z$ is governed by Eq.~(\ref{eq:zEq_1d})
which in turn yields an equation for the dynamics of $J$:
\begin{eqnarray}
\ddot{J}=AJ-\dot{J}\,,\label{eq:EqmJ_1d}
\end{eqnarray}
 with $A=\partial u/\partial x$. This is the one-dimensional
analogue of Eq.~(2.20) in \cite{Ijz10}.

The singularities $z\rightarrow-\infty$ and $J\rightarrow0$ occur
simultaneously. This can be seen in the deterministic limits of Eqs.~(\ref{eq:zEq_1d})
and (\ref{eq:EqmJ_1d}): assume that $z$ is large.
Then (\ref{eq:zEq_1d}) can be approximated by $\dot{z}=-z-z^{2}$.
When $J$ is small then (\ref{eq:EqmJ_1d}) is approximately $\ddot{J}=-\dot{J}$.
These two equations are solved by
\begin{equation}
z=\frac{z_{0}}{(1+z_{0})e^{t}-z_{0}}\,,\hspace{0.5cm}J=J_{0}(1+z_{0}(1-e^{-t}))\,.\label{eq:deterministic_sol_1d}
\end{equation}
Consider an initial condition $z_{0}<-1$. In this case, singularities
in $z$ and $J$ occur as $t$ passes through $t_{0}=\ln(z_{0}/(1+z_{0}))$
for both solutions (\ref{eq:deterministic_sol_1d}). Thus, the rate
at which $J$ passes 0 is identical to the rate at which $z$ tends
to $-\infty$.

\subsection{Two and three spatial dimensions}

In two and three spatial dimensions the situation is analogous. The
matrices $\ma Z$ and $\ma J$ are related by $\ma Z=\dot{\ma J}\ma J^{-1}$.
Eq.~(\ref{eq:zEq}) gives the motion of $\ma Z$ and the corresponding
equation for $\ma J$ is
\begin{equation}
\ddot{\ma J}=\ma A\ma J-\dot{\ma J}\label{eq:EqmJ}\,.
\end{equation}
This equation is identical to Eq.~(2.20) in \cite{Ijz10}.
 In analogy with the one-dimensional case, the deterministic solution
is found to be:
\begin{equation}
\ma J=(1+\ma Z_{0}(1-e^{-t}))\ma J_{0}\,.\label{eq:deterministic_sol}
\end{equation}
 where $\ma Z=\dot{\ma J}\ma J^{-1}$ is obtained from (\ref{eq:deterministic_sol}).
Singularities occur when the determinant $J\equiv\det(\ma J)$ vanishes,
or equivalently when $\tr\ma Z=\dot{J}/J$ diverges. The determinant
of $\ma J$ is obtained from (\ref{eq:deterministic_sol}) in two
and three spatial dimensions
\begin{eqnarray}
J_{d=2} & =&J_{0}[1+T_{1}+Z_{0}-e^{-t}(T_{1}+2Z_{0})+Z_{0}e^{-2t}]\\
J_{d=3} & =&J_{0}[1+T_{1}+T_{2}+Z_{0}-e^{-t}(T_{1}+2T_{2}+3Z_{0}) \nonumber\\&&+e^{-2t}(3Z_{0}+T_{2})-Z_{0}e^{-3t}]\,,
\nonumber
\end{eqnarray}
 where the invariants $J_{0}\equiv\det\ma J_{0}$, $Z_{0}\equiv\det\ma Z_{0}$,
$T_{1}\equiv\tr\ma Z_{0}$ and $T_{2}\equiv[(\tr\ma Z_{0})^{2}-\tr(\ma Z_{0}^{2})]/2$
were defined. Depending on the initial condition $\ma Z_{0}=\dot{\ma J}_{0}\ma J_{0}^{-1}$,
$J$ may pass zero at a finite time $t_{0}$. Now $\dot J$ and $J$ cannot pass zero simultaneously (assuming that $J(t)$ is a regular function, then $\dot J(t_0)=0$ implies that
$J(t)$ has a double root at $t_0$). It follows that $\tr\ma Z$ is singular at $t_0$.
We have explicitly checked in two spatial dimensions that this is
the case.

\section{Conclusions}

\label{sec:conc} In this paper we have compared three recent approaches
to describing inertial particle dynamics: caustic formation giving
rise to multi-valued particle velocities, the notion of random uncorrelated
motion, and spatial clustering as a consequence of singularities in
the local deformation tensor $\ma J$.

We have shown that clustering due to singularities of $\ma J$ can be 
explained in terms of caustic formation. Furthermore we have
compared the consequences of the hypothesis of random uncorrelated
motion with predictions for the fluctuations of relative velocities
in random-flow models. The hypothesis of random uncorrelated motion
leads to an expression for the moments of relative velocities that consists of two terms:
a smooth part, and a contribution due to random uncorrelated motion.
This expression corresponds precisely to Eqs.~(\ref{eq:mp_1d}) and (\ref{eq:mp}) for the moments
of relative velocities obtained in \cite{Gus10}. These theoretical results, describing  
the effect of caustics upon the fluctuations of relative velocities, make it possible
to quantify the degree of random uncorrelated motion, commonly measured
in terms of the longitudinal structure function $s^{(2)}(R)$: for Stokes
numbers below a critical value, $s^{(2)}(R)$ tends to zero as the separation
$R\rightarrow 0$. 

We have performed numerical simulations of one- and two-dimensional
random-flow models in the white-noise limit as well as kinematic
simulations at finite Kubo numbers. We have found that results of these simulations
are consistent with Eqs.~(\ref{eq:mp_1d}) and (\ref{eq:mp}).

Recently, two comprehensive studies of inertial particle dynamics using
direct numerical simulations of particles suspended in turbulent
flows were published \cite{Bec10,Collins12}.
A detailed comparison between the analytical theory
and the results of these direct numerical simulations
for the distribution and the moments of relative
velocities will be published elsewhere \cite{Gus12b}.

Last but not least, we remark that the phenomenon of clustering and relative particle dynamics in turbulent flows analysed here
has much in common with the way particles are transported and deposited in turbulent boundary layers \cite{You97}: enhanced particle concentrations
are observed near the wall, corresponding to the clustering of inertial particles in turbulent flows.
Moreover, as in the case of particles suspended in turbulent flows,
particle inertia gives rise to large impact velocities [referred to as \lq free flight to the wall'
\cite{Broo94}].

{\em Acknowledgments.} We gratefully acknowledge financial
support from Vetenskapsr\aa{}det, from the G\"oran Gustafsson Foundation
for Research in Natural Sciences and Medicine,
and from the EU COST Action MP0806 on "Particles in Turbulence".

\end{document}